\documentclass[%
 reprint,
 amsmath,amssymb,
 aps,
]{revtex4-2}

\usepackage{graphicx} 
\usepackage{dcolumn} 
\usepackage{bm} 
\usepackage{placeins} 

\begin{document}

\preprint{APS/123-QED}

\title{Elastocapillary menisci mediate interaction of neighboring structures at the surface of a compliant solid}





\author{Lebo Molefe}
\author{John M. Kolinski}
\email{john.kolinski@epfl.ch}
\affiliation{
Department of Mechanical Engineering (IGM), School of Engineering (STI), Ecole Polytechnique Fédérale de Lausanne (EPFL), 1015 Lausanne, Switzerland
}

\date{\today}

\begin{abstract}


Surface stress drives long-range elastocapillary interactions at the surface of compliant solids, where it has been observed to mediate interparticle interactions and to alter the transport of liquid drops. We show that such an elastocapillary interaction arises between neighboring structures that are simply protrusions of the compliant solid. For compliant micropillars arranged in a square lattice with spacing $p$ less than an interaction distance $p^*$, the distance of a pillar to its neighbors determines how much it deforms due to surface stress: pillars that are close together tend to be rounder and flatter than those that are far apart. The interaction is mediated by the formation of an elastocapillary meniscus at the base of each pillar, which sets the interaction distance and causes neighboring structures to deform more than those that are relatively isolated. Neighboring pillars also displace toward each other to form clusters, leading to the emergence of pattern formation and ordered domains.
\end{abstract}

\maketitle

Surface stress can completely change the shape of the interface of a compliant solid at sufficiently small scales \cite{surface_tension_solids_1950, on_pattern_transfer_replica_molding_2008, surface_tension_induced_flattening_nearly_plane_elastic_solid_2012, capillarity_driven_instability_soft_solid_2010, solid_drops_large_capillary_deformations_immersed_elastic_rods_2013, surface_folding_induced_attraction_motion_particles_soft_elastic_gel_2013, surface_energy_strained_amorphous_solids_2018, singular_dynamics_failure_soft_adhesive_contacts_2019, elastocapillary_interaction_particles_surfaces_ultrasoft_gels_2014, flattening_patterned_compliant_solid_surface_stress_2014, liquid_drops_attract_repel_inverted_cheerios_2016, elastocapillarity_surface_tension_mechanics_soft_solids_2017, surface_stress_surface_tension_polymeric_networks_2018, statics_dynamics_soft_wetting_2020, elastocapillarity_when_surface_tension_deforms_elastic_solids_2018, extracting_surface_tension_soft_gels_elastocapillary_wave_behavior_2018, elastowetting_soft_hydrogel_spheres_2018, surface_textures_suppress_viscoelastic_braking_soft_substrates_2020, how_surface_stress_transforms_rough_profiles_adhesion_rough_elastic_bodies_2020, surface_tension_strain_dependent_topography_soft_solids_2021, pinning_induced_folding_unfolding_asymmetry_adhesive_creases_2021}, just as it leads to large deformation of small fluid volumes \cite{why_surface_tension_force_parallel_to_interface_2011, experiments_capillary_instability_liquid_jet_1965}. The excess tangential stress in a liquid-air interface promotes the instability of a cylindrical liquid jet, causing it to break into droplets \cite{experiments_capillary_instability_liquid_jet_1965} and surface stress has been observed to cause an analogous instability for cylinders of compliant solid gels \cite{capillarity_driven_instability_soft_solid_2010}. Compliant solids routinely undergo large deformation in response to stresses that they are exposed to. However, the consequences for an interfacial structure that deforms subject to surface stress remain poorly understood.

The typical scale at which elastocapillary phenomena are expected to be observed is defined by the ratio of the surface stress $\Upsilon$ to the material's elastic modulus $E$, and is called the elastocapillary length $\ell_{ec} = \Upsilon / E$ \footnote{In general, $\mathbf{\Upsilon}$ is a tensor, but constant $\Upsilon$ is used in the uniform and isotropic case.}; this arises from a comparison of capillary pressure to the elastic strain in the bulk \cite{elastocapillarity_surface_tension_mechanics_soft_solids_2017}. For structures at a solid-fluid interface with radius of curvature $\mathcal{R}$ below $\ell_{ec}$, surface stress prompts rounding and large deformations \cite{elastocapillarity_surface_tension_mechanics_soft_solids_2017, statics_dynamics_soft_wetting_2020}. In this way, regions of a surface that have initially small radius of curvature, such as the corners of a sharp feature, deform into regions of large radius of curvature that extend farther from the original feature and are in some cases reminscent of liquid menisci. It is unknown how the deformation of a single structure influences the deformation of other structures at the interface, and ultimately the entire interface geometry.

Here we probe the mutual interaction of deforming structures at the interface of compliant gels by molding grids of three-dimensional (3D) cuboid micropillar features with size similar to the material's elastocapillary length, and varying the center-to-center spacing $p$ between the microtextures. Following release from the mold, the solid-air interface deforms due to surface stress. We experimentally study the deformed micropillar surface profiles through two metrics: their final deformed height $h_d$ and mean curvature $H$. An elastocapillary interaction emerges below an interaction distance $p^*$. At distances $p < p^*$, pillar deformation is strongly dependent on the location of neighboring pillars.

The spacing-dependent pillar deformation generates two regimes of the system's behavior: above the interaction distance $p^*$, a pillar's deformation depends only on its initial shape and material properties ($\ell_{ec}$). Below $p^*$, however, proximity to neighboring structures increases deformation. The interaction is shown to be mediated by elastocapillary menisci formed at the base of each pillar. Within the elastocapillary interaction regime, besides the increase of individual pillar deformation, additional behaviors such as clustering emerge. In addition to its relevance for practical applications in microfabrication and other areas \cite{on_pattern_transfer_replica_molding_2008}, our study also improves our fundamental understanding of surface stress, including highlighting the emergence of ordered domains.

\begin{figure}[!ht]
    \centering
    \includegraphics[width=0.46\textwidth]{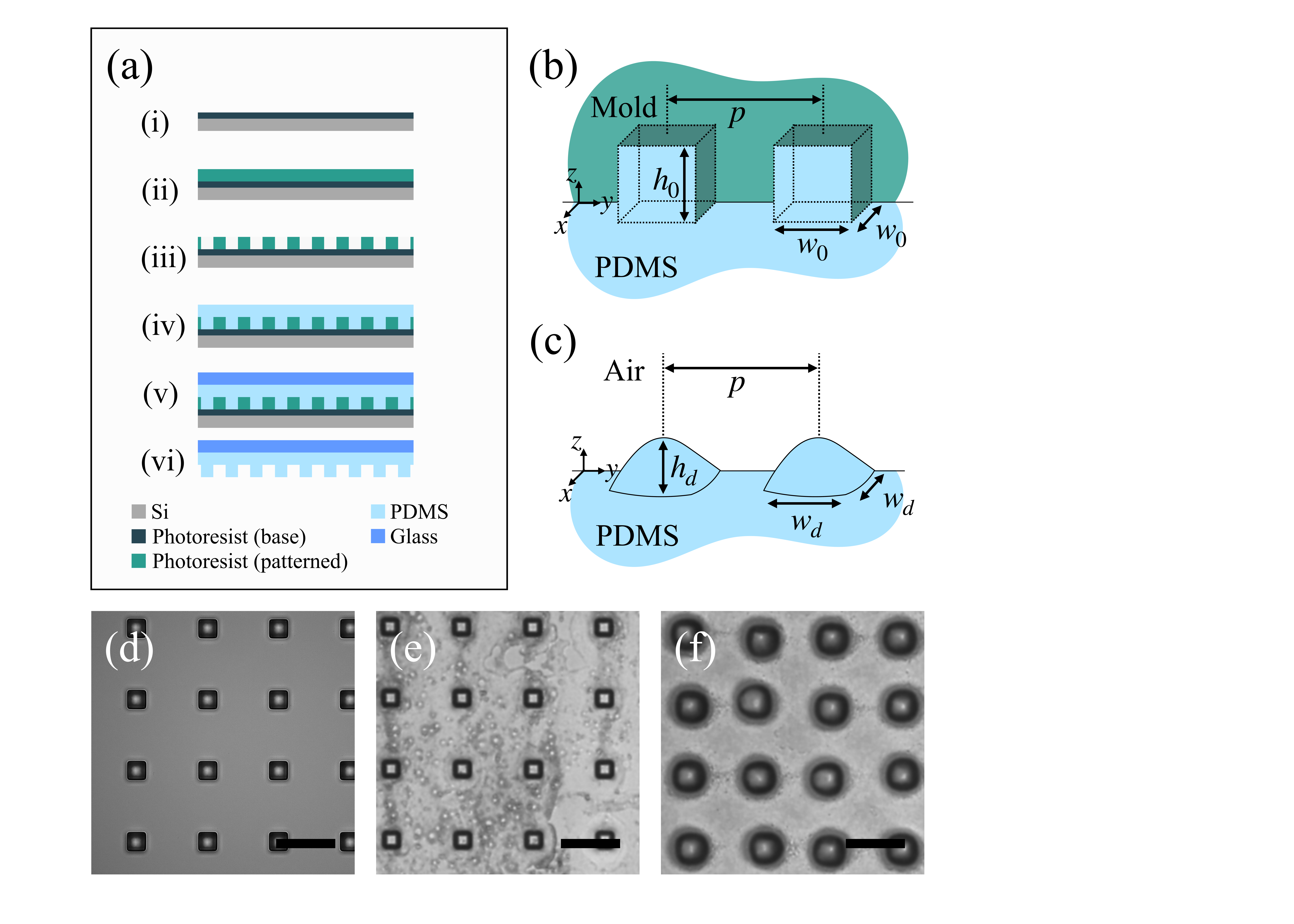}
    \caption{Surface fabrication using a dissolvable mold. (a) Microfabrication procedure: A silicon wafer is coated with \textit{(i)} a 2 \textmu m layer of photoresist (AZnLOF2020, Merck), followed by \textit{(ii)} a $30$ \textmu m layer of photoresist (AZ40XT, Merck). \textit{(iii)} The top layer is patterned using photolithography. \textit{(iv)} An approximately $200$ \textmu m thick layer of compliant gel (PDMS CY 52-276, Dowsil) is spin coated on the mold and cured by baking. \textit{(v)} A glass wafer is bonded to the gel surface using oxygen plasma. \textit{(vi)} Photoresist layers are subsquently dissolved with acetone, after which the surface is immersed in water to rinse, then allowed to dry in ambient conditions; thus, the textured surface is released. (b) The mold shape is a grid of recessed rectangular prisms with center-to-center spacing $p$. (c) Upon release from the mold, the PDMS deforms to height $h_d$ and width $w_d$. (d--f) Optical microcoscopy images for $p = 90$ \textmu m of the: (d) dissolvable mold; (e) stiff surface, $E \approx 190$ kPa; (f) compliant surface, $E \approx 3$ kPa. The mold is well-replicated by the stiff PDMS, whereas significant deformation is observed for the compliant PDMS. Scale bars are 100 \textmu m.}
    \label{fig:microfabrication_procedure_and_surface_images}
\end{figure}

To make textured surfaces in a compliant solid material (E $\lesssim 3$ kPa), we use the microfabrication technique shown in Fig.~\ref{fig:microfabrication_procedure_and_surface_images}(a). Rather than attempting to separate the surface cleanly from a mold by peeling \cite{anti_stiction_coating_PDMS_double_replica_moulding_2011, PDMS_double_casting_method_plasma_treatment_alcohol_passivation_2019, increasing_silicone_mold_longevity_review_PDMS_double_casting_2020, surface_tension_strain_dependent_topography_soft_solids_2021, flattening_patterned_compliant_solid_surface_stress_2014}, where textures are prone to breaking, we instead cast a compliant silicone gel (PDMS CY 52-276, Dow Europe GmbH) into a dissolvable mold [Fig.~\ref{fig:microfabrication_procedure_and_surface_images}(a)(iii--iv)]. The method of using a dissolvable mold has previously been used to study surface stress deformation of millimeter-scale elastic rods using styrofoam molds dissolved in toluene \cite{capillarity_driven_instability_soft_solid_2010, solid_drops_large_capillary_deformations_immersed_elastic_rods_2013} and we demonstrate that this idea can also be applied to produce micrometer-scale patterns. 

First, a mold is produced from two polymeric layers: the first is a sacrificial layer in contact with the tops of the pillars and the second is a photoresist patterned by photolithograpy, which shapes the pillars' side walls [Fig.~\ref{fig:microfabrication_procedure_and_surface_images}(a)(i--iii)]. A schematic of the mold is shown in Fig.~\ref{fig:microfabrication_procedure_and_surface_images}(b), where each hole has depth $h_0 = 26$ \textmu m and width $w_0 = 30$ \textmu m. The holes are arranged on a square lattice with spacing $p$ varying between $45-480$ \textmu m. Fig.~\ref{fig:microfabrication_procedure_and_surface_images}(d) is an image of the mold for $p = 90$ \textmu m. The initial mold dimensions were verified by optical microscopy (for $w_0$) and optical profilometry (for $h_0$). After spin-coating a compliant gel into the mold and curing the material by baking at $80^{\circ}$C for 30 minutes, the gel layer is bonded to a glass surface for support [Fig.~\ref{fig:microfabrication_procedure_and_surface_images}(a)(v)]. The whole ensemble is then placed in an acetone bath, where the mold dissolves, but the cast gel is left intact [Fig.~\ref{fig:microfabrication_procedure_and_surface_images}(a)(vi)] \footnote{Acetone has been shown to swell PDMS by 6\% \cite{solvent_compatibility_PDMS_based_microfluidic_devices_2003}, yet because this is a slight effect, we assume gel deformation is not significantly affected by the solvent, especially since acetone is volatile and does not remain within the gel once it is dry.}. After dissolution of the mold, we remove the surface from acetone, immerse the surface in deionized water to rinse off the acetone, and dry the PDMS by leaving it in air. 

A white light interferometric profilometer (Bruker Contour X) is used to measure the solid-air interface profile. Pillar height $h_0$ is small compared to the bulk thickness, estimated to be 200 \textmu m based on the spin-coating speed. To vary the material stiffness, we vary the ratio of part A to part B in PDMS CY 52-276. A 1:1 A:B ratio produces a compliant surface and a 1:10 A:B ratio produces a stiff surface. A schematic of the released textures is shown in Fig.~\ref{fig:microfabrication_procedure_and_surface_images}(c), where they deform to a final height $h_d$ and width $w_d$. This technique can be used to produce 3D pillar textures [Figs.~\ref{fig:microfabrication_procedure_and_surface_images}(e)--\ref{fig:microfabrication_procedure_and_surface_images}(f)] as well as more complex shapes.

To characterize the elastocapillary length, we measure the gel's shear modulus $G$ with a rheometer (Anton Paar MCR 301). For the compliant material $G = 0.84 \pm 0.07$ kPa and for the stiff material $G = 63 \pm 2$ kPa. The surface stress is $\Upsilon = 35 \pm 5$ mN/m and was measured by imaging the wetting ridge of a glycerol drop deposited on the PDMS with a confocal fluorescence microscope (Leica SP8 STED 3X with a 93$\times$ glycerol immersion lens, NA = 1.3). See the Supplemental Material for fabrication and material characterization details \cite{supplemental_material}.

\begin{figure}[!htbp]
    \centering
    \includegraphics[width=0.46\textwidth]{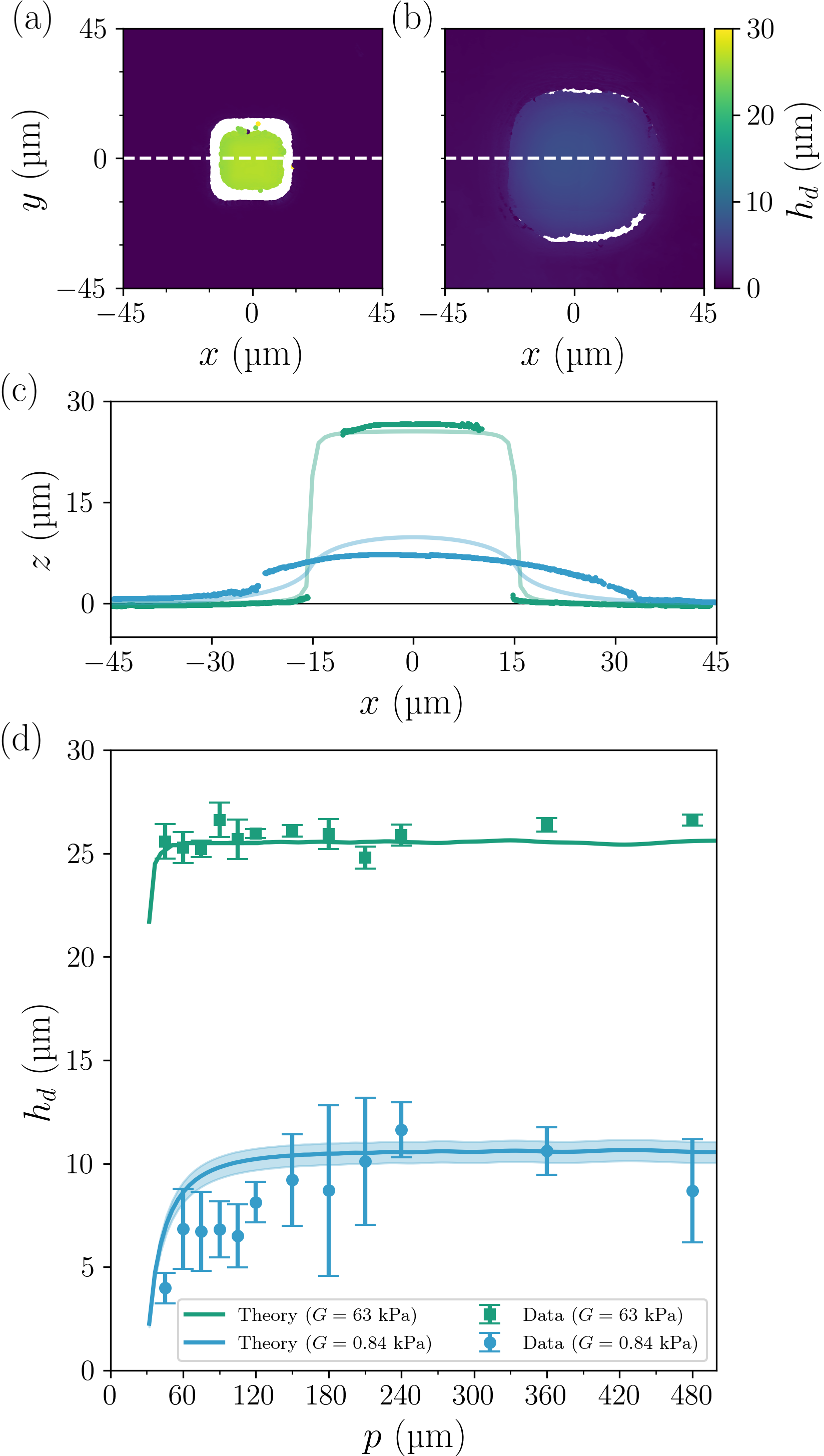}
    \caption{Deformed surface profiles. (a) Height profile for a stiff micropillar, $E \approx 190$ kPa. (b) Height profile for a compliant micropillar, $E \approx 3$ kPa. Grid spacing in both cases is $p = 90$ \textmu m. (c) Profiles of the stiff and compliant pillars as viewed along the dashed lines indicated in (a) and (b). Data are compared to a theoretical model for low-aspect-ratio features~\cite{how_surface_stress_transforms_rough_profiles_adhesion_rough_elastic_bodies_2020} (lines). For the compliant pillar $h_d$ is less than predicted. (d) Deformed height $h_d$ of stiff pillars (squares) and compliant pillars (circles) as a function of grid spacing $p$. For compliant pillars, $h_d$ decreases as $p$ becomes small. The theory (solid lines)~\cite{how_surface_stress_transforms_rough_profiles_adhesion_rough_elastic_bodies_2020} is plotted with a shaded confidence band representing a 95\% confidence interval in the elastic modulus. The confidence band for the stiff material is too small to be visible.}
    \label{fig:deformation_profiles}
\end{figure}

\begin{figure}[!hbtp]
    \centering
    \includegraphics[width=0.46\textwidth]{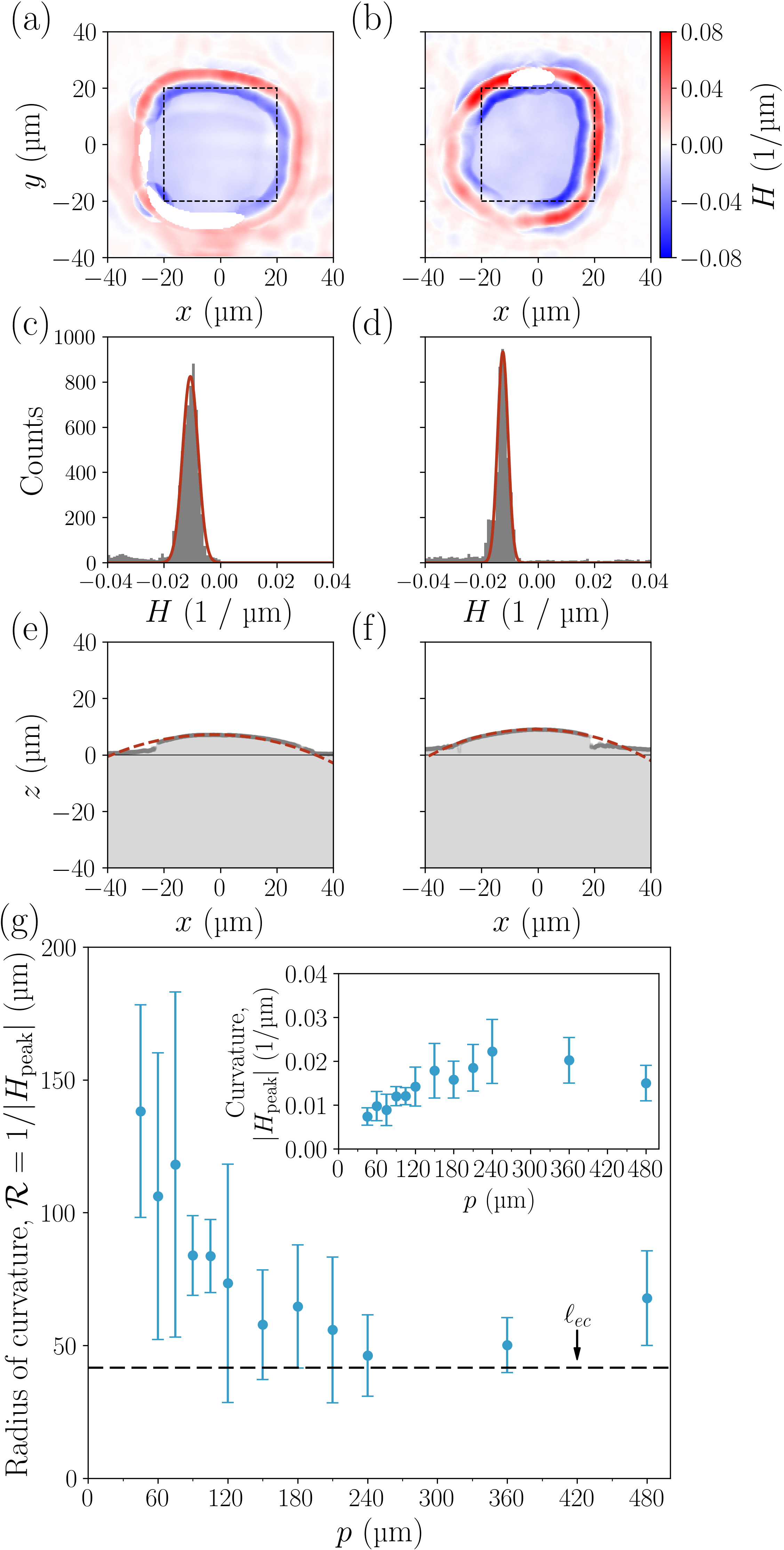}
    \caption{Deformed curvature. (a--b) Mean curvature field $H(x, y)$ for a deformed compliant micropillar ($E \approx 3$ kPa) with spacing (a) $p = 90$ \textmu m and (b) $p = 480$ \textmu m. Negative curvature indicates that the pillar curves toward the $-\hat{z}$ direction. (c) Histogram of $H$ for the pillars with (c) $p = 90$ \textmu m and (d) $p = 480$ \textmu m, considering data inside the dashed boxes in (a--b). Gaussian fits (solid lines) have peaks at $H_{\text{peak}} = -0.010 \pm 0.003$ 1/\textmu m and $H_{\text{peak}} = -0.012 \pm 0.002$ 1/\textmu m, respectively (68\% confidence interval). (e--f) Circular arcs (dashed red lines) with radius of curvature (e) $\mathcal{R} = 94$ \textmu m and (f) $\mathcal{R} = 80$ \textmu m are superimposed on the pillar profiles (gray points). (g) Radius of curvature $\mathcal{R}$ as a function of spacing $p$. As pillars $p \to 0$, radius of curvature increases -- in other words, pillars are flatter. Dashed line is the elastocapillary length $\ell_{ec}$. (inset) Curvature $|H_{\text{peak}}|$ as a function of $p$.}
    \label{fig:curvature}
\end{figure}

\begin{figure}[!htbp]
    \centering
    \includegraphics[scale=0.48]{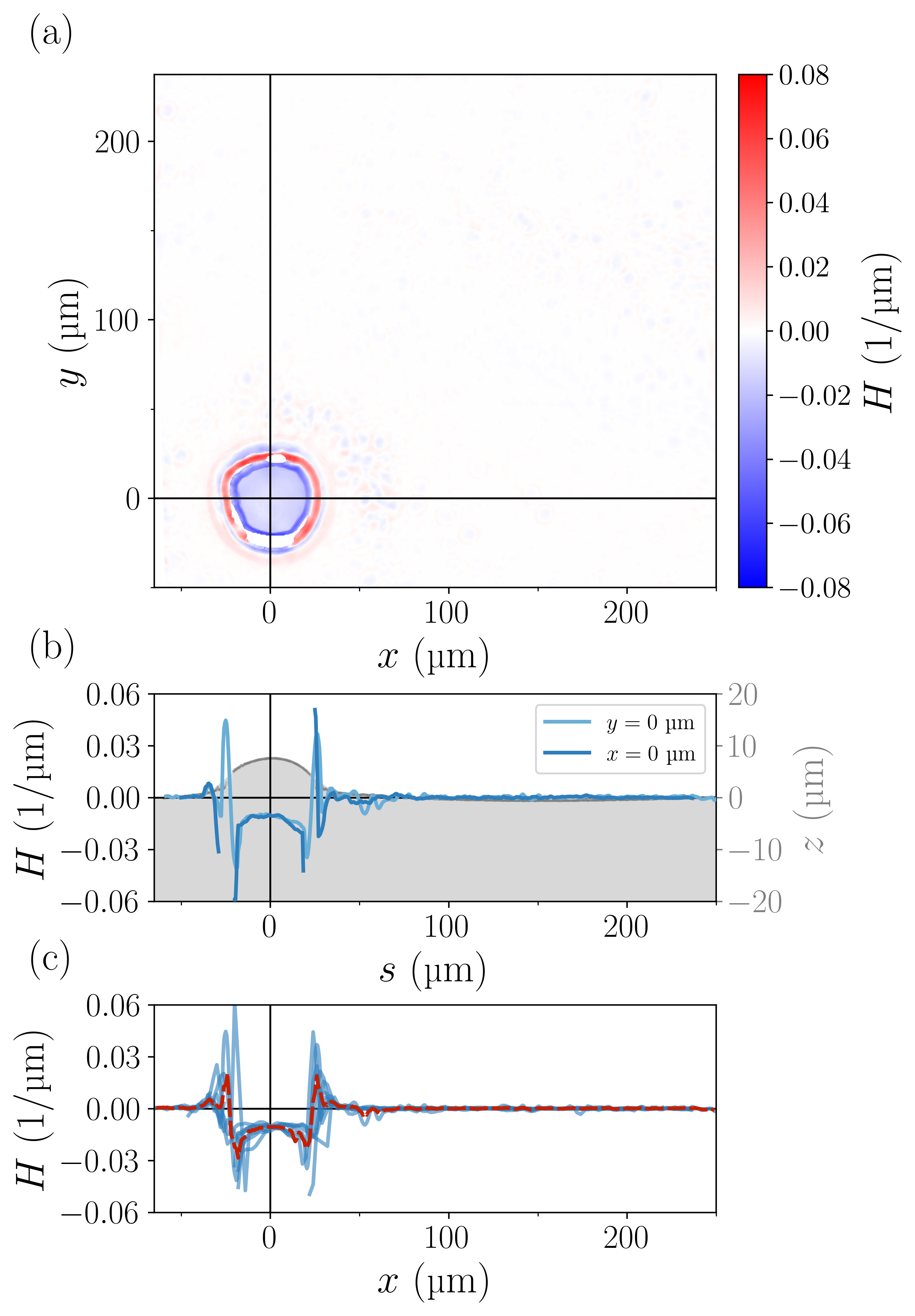}
    \caption{Elastocapillary meniscus scale. (a) Mean curvature field $H(x, y)$ around a pillar with $p = 480$ \textmu m. (b) $H$ as a function of position $s$ along $y = 0$ (light blue) and $x = 0$ (dark blue). The pillar profile is plotted in gray. (c) Curvature along $y = 0$ for 8 different pillars with $p = 480$ \textmu m. The red dashed line is a moving average using bin width of 3 \textmu m. Beyond $\sim$60 \textmu m from the pillar center, curvature decays to zero.}
    \label{fig:curvature_decay}
\end{figure}

\begin{figure*}[!htb]
    \centering
    \includegraphics[scale=0.23]{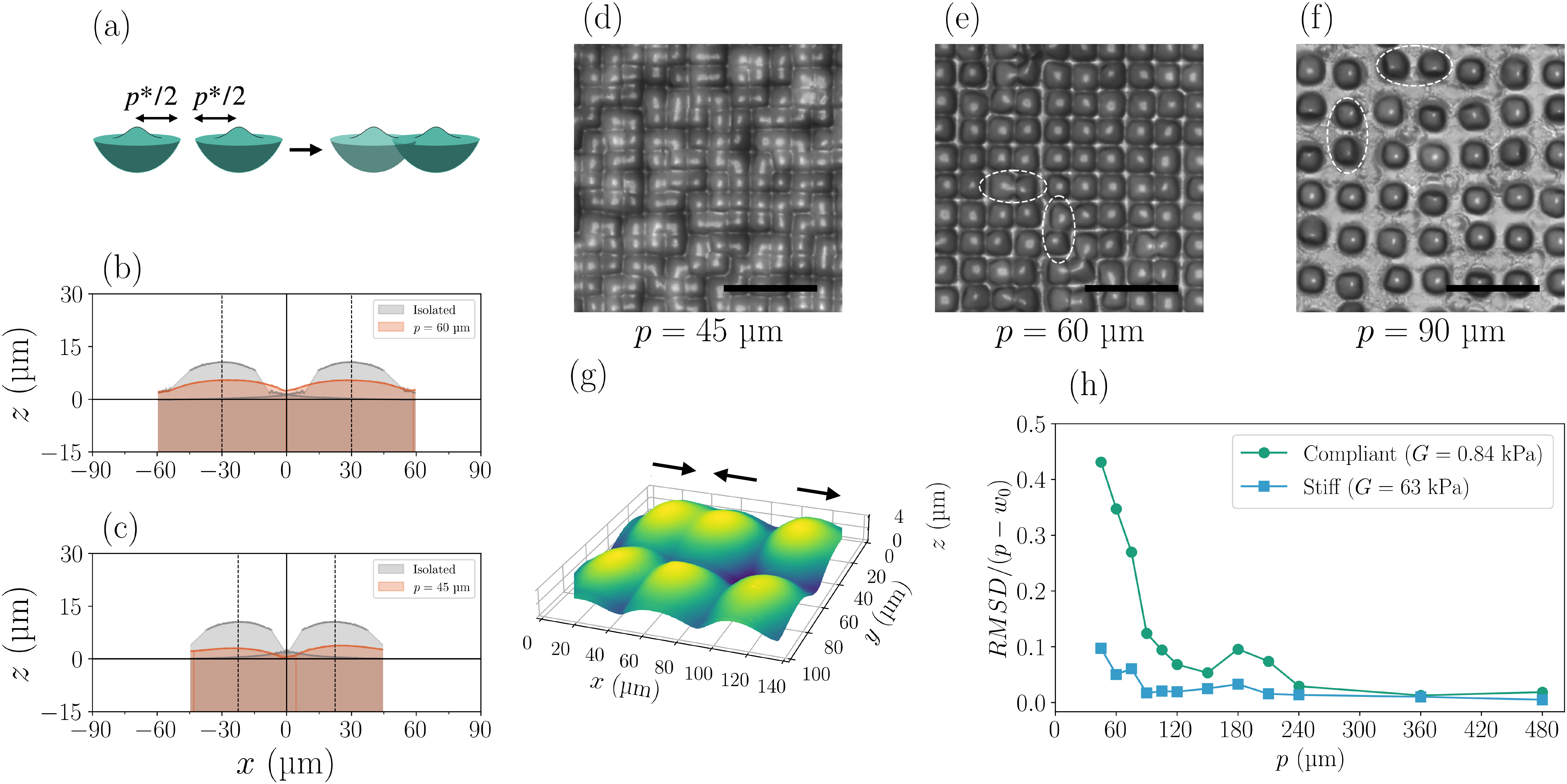}
    \caption{Interaction of elastocapillary menisci. (a) Schematic of a pair of pillars depicting elastocapillary meniscus scale $s^* \approx p^*/2$. When the distance between adjacent pillars is below $p^*$, pillars interact. (b--c) Profiles of adjacent pillars (orange) at spacings: (b) $p = 60$ \textmu m and (c) $p = 45$ \textmu m. Gray profiles in each plot are profiles of a pillar at large spacing ($p = 360$ \textmu m) that is duplicated so we can visualize the hypothetical overlap of its elastocapillary meniscus with another isolated pillar. (d--f) Optical microscopy of compliant pillar clusters at (d) $p = 45$ \textmu m, (e) $p = 60$ \textmu m, and (f) $p = 90$ \textmu m. Scale bars are 200 \textmu m. (g) Profilometry data depict arrested coalescence of a clustered pair for $p = 90$ \textmu m. (h) Clustering behavior, quantified by pillar centroids' root-mean-squared deviation (RMSD), divided by $(p - w_0)$.}
    \label{fig:pillar_interaction}
\end{figure*}

Fig.~\ref{fig:deformation_profiles} compares the deformed profiles of stiff [Fig.~\ref{fig:deformation_profiles}(a)] and compliant [Fig.~\ref{fig:deformation_profiles}(b)] micropillars at $p = 90$ \textmu m. The stiff pillar has replicated the mold with fidelity, having the mold's original height $h_0 = 26$ \textmu m, and also retaining the original lateral dimensions of the mold (30 \textmu m $\times$ 30 \textmu m), whereas the compliant one has become much shorter, rounding into an almost circular profile and expanding significantly in the $xy$-plane. Fig.~\ref{fig:deformation_profiles}(c) displays a side-view of the profiles. We compare the profiles to the predicted deformation in a theory proposed for 3D structures with low aspect ratio proposed by Hui et al. \cite{how_surface_stress_transforms_rough_profiles_adhesion_rough_elastic_bodies_2020} and find that it models the stiff pillar well but overpredicts the height of the compliant one. This suggests that the theory may need to be modified for structures that have a characteristic scale similar to $\ell_{ec}$ but do not satisfy the assumption of infinitesimal aspect ratio. Significant rounding and flattening of compliant pillars were similarly observed for all spacings $p = 45 - 480$ \textmu m.

Fig.~\ref{fig:deformation_profiles}(d) displays the deformed height $h_d$ measured at the pillar center as a function of spacing $p$ for surfaces of different stiffness. Error bars are 95\% confidence intervals, where each data point represents the mean height of $30-146$ pillars. The elastocapillary length of the stiff material is $\ell_{ec} \approx 0.5$ \textmu m, so that we expect surface stress to have a minimal effect on structures at a 30 \textmu m scale. Indeed, the stiff pillar heights match the original mold design ($h_0  = 26$ \textmu m) across all spacings $p$. By contrast, the elastocapillary length of the compliant material is $\ell_{ec} \approx 42$ \textmu m, so we expect surface stress to drive significant deformations. A compliant pillar's height $h_d$ is highly dependent on spacing for low values of $p$, whereas $h_d$ reaches a constant value as $p$ increases, as is evident in Fig.~\ref{fig:deformation_profiles}(d). We identify the length scale for the transition between these two behaviors around $p^* \approx 120-150$ \textmu m, which defines the elastocapillary interaction distance. For $p < p^*$, a pillar's deformation is affected by its neighbors, whereas for $p > p^*$, deformation depends only on the pillar's initial shape and material properties $G$ and $\Upsilon$.

We observe a similar trend in the mean curvature $H$ as a function of $p$ \footnote{Mean curvature $H$ is computed by fitting a paraboloid to local subsets of the pillar surface, from which we compute the curvature at each point, as described in detail in the Supplemental Material \cite{supplemental_material}.}. Curvature fields for pillars at $p = 90$ \textmu m and $p = 480$ \textmu m are shown in Figs.~\ref{fig:curvature}(a--b). For both spacings $p$, the curvature across the top of the pillar is negative, indicating that the pillar curves toward the surface. Moreover, $H$ varies little across the pillar center, as demonstrated by the narrow spread in the histograms of $H$ shown in Figs.~\ref{fig:curvature}(c)--\ref{fig:curvature}(d), indicating that the initial cuboids have deformed into approximately spherical caps with radius $\mathcal{R} = 1/|H_\text{peak}|$, where $H_\text{peak}$ is the value of $H$ at the location of the Gaussian peak. The radius of curvature is $\mathcal{R} = 94$ \textmu m for the pillar with $p = 90$ \textmu m, indicating that it is flatter than the relatively isolated pillar with $p = 480$ \textmu m, for which $\mathcal{R} = 80$ \textmu m. Circular arcs with these $\mathcal{R}$ values overlaid onto the profiles are an excellent match [Figs.~\ref{fig:curvature}(e)--\ref{fig:curvature}(f)]. Fig.~\ref{fig:curvature}(g) displays $\mathcal{R}$ as a function of $p$ for compliant pillars. $\mathcal{R}$ is largest at small spacings, whereas it decreases to a constant value as $p$ increases, indicating that close pillars flatten significantly more (having larger radius of curvature) than relatively isolated ones. Notably, the transition to a constant curvature occurs at the same distance $p^* \approx 120-150$ \textmu m identified in the analysis of $h_d$ [Fig. \ref{fig:deformation_profiles}(d)].

The increased deformation and rounding for $p < p^*$ suggests an elastocapillary interaction between pillars. To probe this interaction, we measure the curvature field $H(x, y)$ around a relatively isolated pillar, as shown in Fig.~\ref{fig:curvature_decay}(a). In Fig.~\ref{fig:curvature_decay}(b), we observe that the curvature field has almost constant negative curvature near the top of the pillar, which becomes positive where the pillar meets the surface, and subsequently decays to zero. By doing the same analysis for several pillars [Fig.~\ref{fig:pillar_interaction}(c)], we observe that the curvature consistently decays with a characteristic distance of $s^* \approx 60$ \textmu m. This horizontal decay length defines the scale of an elastocapillary meniscus. Notably, the pillar shape that was originally square in the $xy$-plane becomes nearly axisymmetric, as is evident in Fig.~\ref{fig:curvature_decay}(a) and discussed in detail in the Supplemental Material \cite{supplemental_material}.

The scale of the curvature decay $s^* \approx 60$ \textmu m is about half of the scale of the interaction distance $p^* \approx 120-150$ \textmu m. This relation $s^* = p^*/2$ is a clear indication that the interaction between pillars is mediated by their elastocapillary menisci: for a pair of pillars, the length scale $2s^*$ is the scale at which the meniscus of one pillar touches that of another pillar -- schematically illustrated in Fig.~\ref{fig:pillar_interaction}(a). This is the same length scale $p^* = 2s^*$ at which we observe spacing-dependent deformation.

To illustrate more clearly what happens to the elastocapillary menisci in the valley between neighboring pillars as compared to isolated ones, in Figs.~\ref{fig:pillar_interaction}(b)--\ref{fig:pillar_interaction}(c) we overlay profiles of two isolated pillars (gray) whose original spacing was $p = 480$ \textmu m as if they were separated by a distance of $p = 60$ \textmu m or $p = 45$ \textmu m. The superposition of two isolated pillars represents a hypothetical scenario where elastocapillary interaction with neighboring pillars does not affect deformation. The region where the menisci overlap near $x = 0$ \textmu m in this hypothetical situation has a higher curvature than the tops of the pillars, and we observe that the profile of a real interface (orange) is rounder by comparison. Thus, as a result of rounding and lifting bulk material between pillars, as well as decreasing pillar height, the meniscus decreases the surface area and curvature, likely reducing the total energy of the system.

Elastocapillary interaction not only increases individual pillar deformation, but also leads to displacement of the pillars' centroids, such that groups of pillars form clusters [Figs. \ref{fig:pillar_interaction}(d)--\ref{fig:pillar_interaction}(f)]. The profile in Fig.~\ref{fig:pillar_interaction}(g) depicts a pair of pillars displaced toward each other. Clustering for a group of pillars can be quantified by the root-mean-squared deviation (RMSD) of pillar centroids from their ideal grid positions, divided by the initial space $(p - w_0)$ between two pillar edges. If the deviation of a pillar and its neighbor toward each other is equal to half the distance between two pillar edges ($\frac{p-w_0}{2}$), the pillars will touch. In Fig.~\ref{fig:pillar_interaction}(f), we show that the RMSD of the ensemble of compliant pillars has a value such that $\text{RMSD}/(p - w_0) \approx 1/2$ for the lowest spacings $p = 45-75$ \textmu m, indicating high levels of clustering that are not observed for the stiff material.

In this work, we demonstrate that neighboring structures at a solid-air interface can interact through an elastocapillary meniscus, leading to pronounced deformation and rounding. We introduce a microfabrication technique for producing structures at the surface of a compliant solid and study how their deformation to final height $h_d$ and mean curvature $H$ depends on grid spacing $p$. Deformation depends on spacing for $p < p^*$ where $p^* \approx 2 s^*$ is set by the horizontal scale $s^*$ of the elastocapillary meniscus around a deformed pillar. Excess deformation due to elastocapillary interaction is significant -- the most closely spaced pillars shrink to around 20\% of their initial height, $h_d \approx 0.2 h_0$. This is much less than the deformed height of a relatively isolated pillar, $h_d \approx 0.5 h_0$.

In addition to large deformations, pillar centroids displace from their initial grid positions, similar to droplet `durotaxis' or the attraction between glass beads at a gel-air interface, which are also mediated by regions of nonzero curvature between the objects \cite{liquid_drops_attract_repel_inverted_cheerios_2016, surface_folding_induced_attraction_motion_particles_soft_elastic_gel_2013, elastocapillary_interaction_particles_surfaces_ultrasoft_gels_2014}. Here we observe a similar long-range elastocapillary interaction between structures that are part of the interface itself, without introducing external objects. This system is an attractive alternative for studying elastocapillary interaction at interfaces, because the singularity at the contact line is not present. Interaction between pillars also differs from the interactions between external objects, however: first, a pillar's displacement toward its neighbor is limited due to the elastic energy cost. Second, the pillar's final shape is affected by the interaction, which does not occur for objects such as rigid beads. Excess deformation and rounding are added mechanisms by which compliant pillars decrease the surface area of their elastocapillary menisci. Intriguingly, we observe the emergence of pattern formation in a system driven solely by the solid's surface stress; by contrast, clustering patterns previously observed for stiff micropillar arrays are driven by liquid-air surface tension \cite{mechanism_resist_pattern_collapse_development_process_1993, capillarity_driven_assembly_two_dimensional_cellular_carbon_nanotube_foams_2004, control_shape_size_nanopillar_assembly_adhesion_mediated_elastocapillary_interaction_2010, two_parameter_sequential_adsorption_model_applied_microfiber_clustering_2010, elastocapillary_coalescence_plates_pillars_2014}. Our observations suggest future directions for introducing defects or varied initial conditions to better characterize emergence of order at low $p$.

Beyond the fundamental study of surface stress, our experiments suggest new directions toward probing self-assembly on gel and elastomer interfaces \cite{surface_folding_induced_attraction_motion_particles_soft_elastic_gel_2013, elastocapillary_interaction_particles_surfaces_ultrasoft_gels_2014, reflow_transfer_conformal_three_dimensional_microprinting_2022} that may open new avenues for microfabrication techniques. Future study of the dynamic response of solid surfaces to changing outer fluids may also lead to new directions for producing soft machines, valves, or other tools to be used, for example, in microfluidic devices.

\begin{acknowledgments}
We thank Guillermo Villanueva, Mojtaba Abdolkhani, Abigail Plummer, and Andrej Košmrlj for insightful discussions. We gratefully acknowledge the staff of the Center for MicroNanotechnology (CMi) at EPFL who provided valuable technical support for the fabrication, in particular Cyrille Hibert, Joffrey Pernollet, Niccolò Piacentini, Georges-André Racine, Gatera Kumuntu, and Vivigan Sinnathamby. Data supporting this study are openly available from \cite{source_data}.
\end{acknowledgments}



\end{document}